\documentclass[preprint,showpacs,preprintnumbers,amsmath,amssymb]{revtex4}

\usepackage{graphicx}%
\usepackage{dcolumn}%
\usepackage{bm}%

\begin{document}

\title{Zipf's Law in Gene Expression} 

\author{Chikara Furusawa}
\affiliation{Center for Developmental Biology, 
The Institute of Physical and Chemical Research (RIKEN), 
Kobe 650-0047, JAPAN
}

\author{Kunihiko Kaneko}
\affiliation{Department of Pure and Applied Sciences
Univ. of Tokyo, Komaba, Meguro-ku, Tokyo 153-8902, JAPAN
}

\date{\today}
             
\begin{abstract}

Using data from gene expression databases on various organisms and tissues,
including yeast, nematodes, human normal and cancer tissues, and embryonic stem cells,
we found that
the abundances of expressed genes exhibit a power-law 
distribution with an exponent close to -1, i.e., they obey Zipf's law.
Furthermore, by simulations of a simple model with an intra-cellular reaction network, 
we found that Zipf's law of
chemical abundance is a universal feature of cells where such a network
optimizes the  efficiency  and faithfulness of  self-reproduction.
These findings provide novel insights into the nature of the organization of 
reaction dynamics in living cells.

\end{abstract}


\maketitle

In a cell, an enormous number of organized chemical reactions are required 
to maintain its living state.
Although enumeration of detailed cellular processes and the construction of
complicated models is important for a complete description of cellular behavior,
it is also necessary to search for universal laws with regard to the
intra-cellular reactions common to all living systems, and then to
unravel the logic of life leading to such universal features.
For example, scale-free networks have recently been discussed as
a universal property of some biochemical reaction networks within existing
organisms \cite{metabolic, lethal}. 
These studies, however, only focused on the properties of the network topologies,
while the reaction dynamics of the networks were not discussed.
Here, we report a universal property of the reaction dynamics that occurs
within cells, namely a  power-law distribution of the
abundance of expressed genes with an exponent close to -1, i.e. a power-law
distribution that obeys Zipf's law \cite{Zipf}.
By using an abstract model of a cell with simple reaction
dynamics, we show that this power-law behavior in the chemical
abundances generally appears when the reaction dynamics leads to 
a faithful and efficient self-reproduction of a cell.
These findings provide insights into the nature of the organization of
complex reaction dynamics in living cells.

In order to investigate possible universal properties of the reaction
dynamics, we examined the distributions of the abundances of expressed
genes (that are approximately equal to the abundances of the
corresponding proteins) in 6 organisms and more than 40 tissues based
on data publicly available from SAGE (Serial Analysis of Gene
Expression) databases \cite{SAGE1, SAGE2, SAGE3}.
SAGE allows the number of copies of any given  mRNA to be
quantitatively evaluated by determining the abundances of the short
sequence tags which uniquely identify it \cite{SAGE_method}.

In Fig.1, we show the rank-ordered frequency distributions of the
expressed genes, where the ordinate indicates the frequency of the observed
sequence tags (i.e. the population ratio of the
corresponding mRNA to the total mRNA), and the abscissa shows the rank
determined from this frequency.
As shown, the distributions follow a power-law with an exponent close to
-1 (Zipf's law).
We observed this power-law distribution for all the available samples,
including 18 human normal tissues, human cancer tissues, mouse 
(including  embryonic stem cells), rat,
nematode ({\it C. elegans}), and yeast ({\it S.cerevisiae}) cells.
All the data over 40 samples (except for 2 plant data) show 
the power-law distributions with the exponent in the range from 
$-1 \sim -0.86$.
Even though there are some factors which may bias the results of the SAGE
experiments, such as sequencing errors and non-uniqueness of tag
sequences, it seems rather unlikely that the distribution is an artifact
of the experimental procedure.

The abundance of each protein is the result of a complex network of chemical reactions
that is
influenced by possibly a large number of factors including other proteins and genes.
Then, why is  Zipf's law universally observed, and
what class of reaction dynamics
will show the observed power-law distribution?
Because the power-law distribution applies to a wide range of existing
organisms, it is expected that
it appears as a general feature of the reaction dynamics of cellular systems.

In order to investigate the above questions, we
adopt a simple model of cellular dynamics
that captures only its basic features. It consists of
intra-cellular catalytic reaction networks that transform nutrient chemicals
into proteins.
By studying a class of simple models with these features, we clarify
the conditions under which the reaction dynamics leads to a power-law
distribution of the chemical abundances.

Of course, real intra-cellular processes are much more 
complicated, but if the mechanism is universal,
the power-law should be valid regardless of how complicated the actual processes are.
Hence it is relevant to study as simple as possible a model when
trying to understand a universal law in real data.

Consider a cell consisting of a variety of chemicals.  The internal state of 
the  cell can be represented
by a set of numbers $(n_1,n_2,\cdots ,n_k)$, where
$n_i$ is the number of molecules of the chemical species $i$ with $i$
ranging from $i=1$ to $k$.
For the internal chemical reaction dynamics, we chose a catalytic
network among these $k$ chemical species, where each reaction from some
chemical $i$ to some other chemical $j$ is assumed to be catalyzed by a
third chemical $\ell$, i.e. $(i + \ell \rightarrow j + \ell)$ .
The rate of increase of $n_j$ (and decrease of $n_i$)
through this reaction is given by 
$\epsilon n_i n_{\ell}/N^2$, where $\epsilon$ is the coefficient for the chemical reaction.
For simplicity all the reaction coefficients were chosen to be equal \cite{hetero}, 
and the connection paths of this catalytic network were chosen randomly such that
the probability of any two chemicals $i$ and $j$ to be connected is given by the connection
rate $\rho$ \cite{stem}.

Some resources (nutrients) are supplied from the environment by diffusion
through the membrane (with a diffusion coefficient $D$), to ensure the growth of a cell.
Through the calaytic reactions, these
nutrients\cite{note1} are transformed into other chemicals.
Some of these chemicals may penetrate \cite{hetero} the membrane 
and diffuse out while others will not.  
With the synthesis of the unpenetrable
chemicals that do not diffuse out, the total number
of chemicals $N= \sum_i n_i$ in a cell can increase, and accordingly
the cell volume will increase.
We study how this cell growth is sustained
by dividing a cell into two when the volume is larger than
some threshold.  For simplicity 
the division is assumed to occur when the total number of molecules $N= \sum_i n_i$ in a cell
 exceeds a given threshold $N_{max}$. Chosen randomly, the mother cell's
molecules are evenly split among the two daughter cells.

In our numerical simulations, we randomly pick up a pair of molecules in a cell, and
transform them  according to the reaction network. 
In the same way, diffusion through the membrane 
is also computed by randomly choosing molecules inside the cell and
nutrients in the environment.   
In the case with $ N \gg k$ (i.e. continuous limit), the reaction dynamics
is represented by the following rate equation:
\begin{center}
\begin{eqnarray}
dn_i/dt  = \sum_{j,\ell}Con(j ,i ,\ell) \;\epsilon \;n_j \;n_{\ell} /N^2~~~~~~~~~ \nonumber \\
- \sum_{j',{\ell}'}Con(i ,j' ,{\ell}') \;\epsilon \;n_i \; n_{{\ell}'} /N^2 
+ D \sigma_i (\overline{n_i}/V - n_i/N), \nonumber
\end{eqnarray}
\end{center}
where $Con(i,j,\ell)$ is 1 
if there is a reaction $i+\ell \rightarrow j + \ell$, and 0 otherwise,
whereas $\sigma_i$ takes 1 if the chemical $i$ is penetrable,
and 0 otherwise.
The third term describes the transport of chemicals through the membrane, 
where $\overline{n_i}$ is a constant, representing  the number 
of the {\it i}-th chemical species 
in the environment
and $V$ denotes the volume of the environment in units of the initial cell size.
The number $\overline{n_i}$ is nonzero only for 
the nutrient chemicals.

If the total number
of molecules $N_{max}$ is larger than the number of chemical species $k$,
the population ratios  $\{ n_i/N \}$  are generally 
fixed, since the daughter cells inherit the  chemical compositions of 
their mother cells.
For $k>N_{max}$ \cite{large_K}, the population ratios do not
settle down and can change from generation to generation.
In both cases, 
depending on the membrane diffusion coefficient $D$,
the intra-cellular reaction dynamics can be classified into
the three classes \cite{rho}. 

First, there is a critical value $D = D_c$ beyond which the cell cannot grow continuously.
When $D > D_c$, the flow of nutrients from the environment is so fast that 
the internal reactions transforming them into chemicals sustaining
`metabolism' cannot keep up.
In this case all the molecules in the cell will finally be substituted by the
nutrient chemicals and the cell stops growing since the nutrients alone cannot catalyze 
any reactions to generate unpenetrable chemicals.
Continuous cellular growth and successive divisions are possible only for $D \leq D_c$.
When the diffusion coefficient $D$ is sufficiently small, the internal reactions
progress faster than the flow of nutrients from the environment, and
all the existing chemical species have small numbers of approximately the same level.
A stable reaction network organization is obtained only at the
intermediate diffusion coefficient below $D_c$, 
where some chemical species have much larger number of molecules 
than others.

The rank-ordered number distributions of chemical species in our model
are plotted
in Fig.2, where the ordinate indicates the number of molecules $n_i$ and
abscissa shows the rank determined by $n_i$.
As shown in the figure, the slope in the rank-ordered number
distribution increases with an increase of the diffusion coefficient $D$.
We found that at the critical point $D = D_c$, 
the distribution converges to a power-law with an exponent -1.

The power-law distribution at this critical point is maintained by a
hierarchical organization of catalytic reactions, where  
the synthesis of higher ranking chemicals is catalyzed 
by lower ranking chemicals.
For example, major chemical species (with e.g. $n_i>1000$) are directly synthesized
from nutrients and catalyzed by chemicals that are slightly less abundant (e.g. $n_i \sim 200$).
The latter  chemicals are mostly synthesized from nutrients 
(or other major chemicals), 
and catalyzed by chemicals that are much less abundant. In turn these  chemicals are  catalyzed
by chemicals that are even less abundant, and this
hierarchy of catalytic reactions continues until it
reaches the minor chemical species (with e.g. $n_i < 5$) \cite{hierarchy}.

Based on this catalytic hierarchy, the observed exponent -1 can be
explained using a mean field approximation.
First, we replace the concentration $n_i/N$ of each chemical $i$,
except the nutrient chemicals, by a single average concentration (mean
field) $x$, while the concentrations of nutrient chemicals $S$ is
given by the average concentration $S=1- k^{*}x$, where $k^{*}$ is the
number of non-nutrient chemical species.
From this mean field equation, we obtain $S=\frac{DS_0}{D+\epsilon
\rho}$ with $S_0=\sum_j \overline{n_j}/{V}$.
With linear stability analysis, the solution with $S \neq 1$ is
stable if $D< \frac{\epsilon \rho}{S_0-1} \equiv D_c$.
Indeed, this critical value does not differ much from 
numerical observation.

Next, we study how the concentrations of non-nutrient chemicals differentiate.
Suppose that chemicals $\{ i_0 \}$ are synthesized directly from 
nutrients through catalyzation by chemicals ${j}$.
As the next step of the mean-field approximation we assume the
concentrations of the chemicals $\{ i_0 \}$ are larger than the
others.
Now we represent the dynamics by two mean-field concentrations; the
concentration of $\{ i_0 \}$ chemicals, $x_0$, and the concentration
of the others, $x_1$.
The solution with $x_0 \neq x_1$ satisfies $x_0 \approx x_1/\rho$ at
the critical point $D_c$.
Since the fraction of the $\{ i_0 \}$ chemicals among the non-nutrient
chemicals is $\rho$, the relative abundance of the chemicals $\{ i_0
\}$ is inversely proportional to this fraction.
Similarly, one can compute the relative abundances of the
chemicals of the next layer synthesized from $i_0$. 
At $D \approx D_c$, this hierarchy of the catalytic 
network is continued.
In general a given layer of the hierarchy is defined by the chemicals
whose synthesis from the nutrients is catalyzed by the layer one step
down in the hierarchy. The abundance of chemical species in a given
layer is $1/\rho$ times larger than chemicals in the layer one step
down.
Then, in the same way as this hierarchical organization of chemicals,
the increase of chemical abundances and the decrease of number of
chemical species are given by factors of $1/\rho$ and $\rho$,
respectively.
This is the reason for the emergence of power-law with
an exponent -1 in the rank-ordered distribution \cite{details}.

In general, as the flow of nutrients from the environment increases,
the hierarchical catalyzation network pops up from random reaction
networks.
This hierarchy
continues until it covers all chemicals, at $D \rightarrow D_c-0$.
Hence, the emergence of a power-law distribution
of chemical abundances near the critical point is quite general, and
does not rely on the
details of our model, such as the network configuration or the kinetic rules
of the reactions.  
Instead it is a universal property of a cell with an intra-cellular
reaction network to grow, by taking in nutrients, at the critical
state, as has been confirmed from simulations of a variety of models.

There are two reasons to assume that such a critical state of the reaction 
dynamics is adopted in existing cellular systems.
First, as shown in Fig.3, the growth speed of a cell is maximal at $D = D_c$.
This suggests that a cell whose reaction dynamics are in  the critical state  
should be selected by natural selection.
Second, at the critical point, the similarity of chemical compositions 
between the mother and daughter cell is maximal as shown in Fig.3.
Indeed, for $k>N$, the chemical compositions differ significantly from generation
to generation when $D \ll D_c$.
When $D \approx D_c$, several semi-stable states with distinct
chemical compositions appear. Daughter cells in the semi-stable states inherit
 chemical compositions that are nearly identical to their mother cells over many 
generations, until
fluctuations in molecule numbers induce a transition to another 
semi-stable state.
This means that the most faithful transfer of the information 
determining a cell's  
intra-cellular state is at the critical state. 
(Inheritance of chemical compositions is  also discussed in \cite{Lancet}
in connection with the origin of  reproducing cells).
In this state, cells of specific chemical compositions are reproduced
and can also 'evolve' into other states.
For these reasons, it is natural to conclude that evolution favors a 
critical state \cite{SOC} for the reaction dynamics. 

Last, we investigated the relationship between the abundance of
a chemical species and the number of reaction paths connected with it.
By comparing the SAGE data and the protein-protein
interaction data in yeast ({\it S.cerevisiae}) 
\cite{yeast_data1, yeast_data2}, 
obtained by systematic two-hybrid analysis, 
we found that there is a significant negative
correlation between the abundance of any given mRNA and the number of
protein-protein interaction links that the corresponding protein takes part
in ($p<0.01$; determined by randomization test).
In our model simulations, this negative correlation between the
abundance of chemical species and the number of possible catalytic
paths of the chemical is also found.
In this sense, chemicals minor in abundance can play a relatively
important role in the control of the behavior of a cell\cite{minority}.
In the future it will be important to
study this kind of interplay in the context of evolution since 
the evolution of reaction networks has only been discussed in 
the context of network topology \cite{metabolic, lethal}.

We would like to thank Tetsuya Yomo and Lars Martin Jakt 
for stimulating discussions and 
Frederick H. Willeboordse and Adam Ponzi for 
critical reading of the manuscript.
Grant-in-Aids for Scientific Research from
the Ministry of Education, Science and Culture of Japan
(11CE2006).

\begin{figure}[htbp]
\begin{center}
\includegraphics[width=11cm,height=15cm]{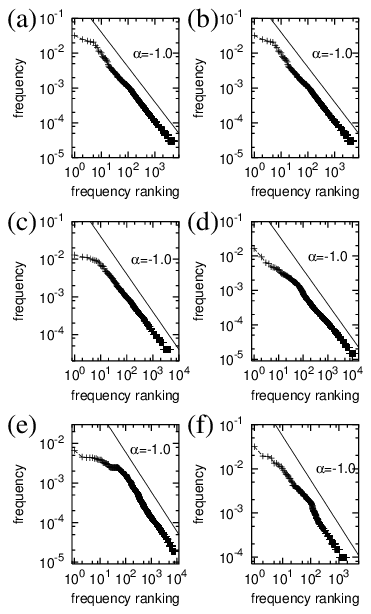}
\end{center}
\caption{
Rank-ordered frequency distributions of expressed genes.
(a), Human liver; (b), kidney; (c), Human colorectal cancer (caco2);
(d), Mouse embryonic stem cells; (e),{\it C. elegans}; 
(f) yeast ({\it Saccharomyces cerevisiae}).
The exponent $\alpha$ of the power law is in the range from $-1 \sim -0.86$ for all
the samples inspected, except for two plant data 
({seedlings of \it Arabidopsis thaliana} and the trunk of {\it Pinus taeda}), 
whose exponents are approximately $-0.63$.
}
\end{figure}

\begin{figure}[htbp]
\begin{center}
\includegraphics[width=11cm,height=9cm]{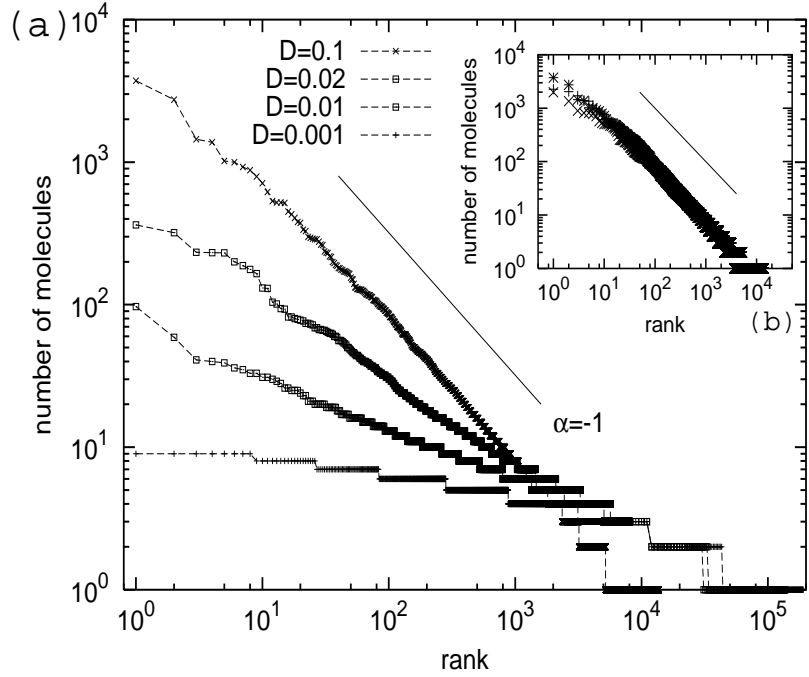}
\end{center}
\caption{
Rank-ordered number distributions of chemical species. 
(a) Distributions with different diffusion coefficients $D$ are overlaid.  The
parameters were set as $k = 5\times 10^6$, $N_{max} = 5\times 10^5$, and $\rho = 0.022$. 
30 \% of chemical species are penetrating the membrane, and others are not.
Within the penetrable chemicals, 10 chemical species are continuously supplied 
to the environment, as nutrients.
In this figure, the numbers of nutrient chemicals in a cell are not plotted.
With these parameters, $D_c$ is approximately $0.1$.
(b) Distributions at the critical points with different total number of
chemicals $k$ are overlaid.
The numbers of chemicals were set as $k=5\times 10^4$, $k=5\times 10^5$, 
and $k=5\times 10^6$, respectively.
Other parameters were set the same as those in (a).
}
\end{figure}

\begin{figure}[htbp]
\begin{center}
\includegraphics[width=11cm,height=9cm]{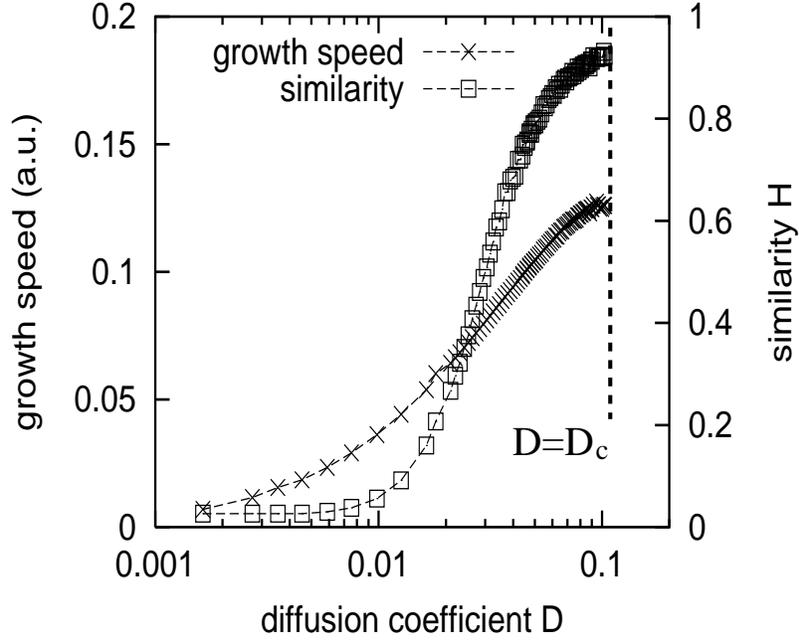}
\end{center}
\caption{
The growth speed of a cell and the similarity between the
chemical compositions of the mother and daughter cells, plotted as a function of
the diffusion coefficient $D$.
The growth speed is measured as the inverse of the time for a cell to divide.
The degree of similarity between two different states $m$ (mother) and $d$ (daughter) 
is measured as the scalar product of k-dimensional vectors
$ H({\bf n}_m,{\bf n}_d) = ({\bf n}_m/|{\bf n}_m|)\cdot ({\bf n}_d/|{\bf n}_d|)$, 
where ${\bf n}=(n_1,n_2,...,n_k)$ represents the chemical composition of a cell 
and $|{\bf n}|$ is the norm of ${\bf n}$ \cite{Lancet}.
Both the growth speed and the similarity are averaged over 500 cell divisions.
Note that the case $H = 1$ indicates an identical chemical composition between the 
mother and daughter cells.
}
\end{figure}


\begin{thebibliography}{999}
\bibitem{hetero}
Even if the reaction coefficient and diffusion coefficient of penetrating chemicals
are not identical but distributed, the results reported here
are obtained.
\bibitem{note1}
The nutrient chemicals have no catalytic activity in order to prevent the
occurrence of catalytic reactions in the environment.
\bibitem{large_K}
Note that, in the case $k>N_{max}$ the number of some chemical species $n_i$ 
is 0, while a sub-population of chemical species sustains the intra-cellular 
dynamics.
\bibitem{rho}
These three classes of intra-cellular dynamics also appear when changing the
 connection rate $\rho$.
There is a critical value $\rho = \rho_c$, where in the case $\rho < \rho_c$ 
the cell stops growing.
The power-law distribution of chemical abundances with an exponent -1 
appears at $\rho = \rho_c$.
\bibitem{hierarchy}
In the case depicted in Fig.2, a hierarchical organization of
catalytic reactions with $5 \sim 6$ layers is observed at the critical point.
\end{thebibliography}
\end{document}